\documentclass[conference]{IEEEtran}
\IEEEoverridecommandlockouts
\usepackage{cite}
\usepackage{amsmath,amssymb,amsfonts}
\usepackage{algorithmic}
\usepackage{graphicx}
\usepackage{textcomp}
\usepackage{xcolor}
\usepackage{mathtools}
\usepackage{commath}

\title{Frequency Control and Disturbance Containment Using Grid-Forming Embedded Storage Networks 
\thanks{The Pacific Northwest National Laboratory is operated for the U.S. Department of Energy by Battelle Memorial Institute under Contract DE-AC05-76RL01830. This research was supported by the Embedded Storage Project funded by the US DOE Office of Electricity (OE).}
}

\author{\IEEEauthorblockN{Kaustav Chatterjee, Ramij Raja Hossain, Sai Pushpak Nandanoori, Soumya Kundu, \\Diane Baldwin, and Ronald Melton }
\IEEEauthorblockA{
\textit{Pacific Northwest National Laboratory}\\
Richland, WA, USA }
}

\begin{document}
\maketitle

\begin{abstract}

The paper discusses fast frequency control in bulk power systems using embedded networks of grid-forming energy storage resources. Differing from their traditional roles of regulating reserves, the storage resources in this work operate as fast-acting grid assets shaping transient dynamics. The storage resources in the network are autonomously controlled using local measurements for distributed frequency support during disturbance events. Further, the grid-forming inverter systems interfacing with the storage resources, are augmented with fast-acting safety controls designed to contain frequency transients within a prescribed tolerance band. The control action, derived from the storage network, improves the frequency nadirs in the system and prevents the severity of a disturbance from propagating far from the source. The paper also presents sensitivity studies to evaluate the impacts of storage capacity and inverter controller parameters on the dynamic performance of frequency control and disturbance localization. The performance of the safety-constrained grid-forming control is also compared with the more common grid-following control. The results are illustrated through case studies on an IEEE test system.
\end{abstract}


\section{Introduction}
In power systems with high renewable penetration, battery energy storage systems (BESS) are important in managing volatility from generation intermittency. Depending on the operating requirements BESS can switch from being a load to a generation resource, and vice-versa. This operational flexibility enhances grid reliability and stability \cite{review_flexibility}. Further, with the recent advances in power electronics, the response times of the BESS can be sufficiently fast to support grid operations at time scales of seconds or even milliseconds \cite{gfm_review, rebalancing}. That said, most grid operators and utilities, for legacy reasons, still view bulk storage systems either as steady-state reserves or as intermediate means for postponing infrastructure investments \cite{review_ibr_support_func}. This restrictive outlook on storage assets leaves most of their potential untapped. 

Historically in a traditional power system, fast-acting grid functions like inertial support, fast frequency response (FFR), and primary frequency response (PFR) have been provided by synchronous generators and their governor-excitation systems \cite{WETO_Gridservice}. However, with the gradual retirement of fossil fuel-fired generators, these support functions in the coming years would have to be sourced from inverter-based resources (IBRs), including BESS \cite{WETO_Gridservice}. To this end, there is a growing interest among developers and IBR manufacturers, to explore appropriate control designs that enable BESS's participation in these applications \cite{distr_control_siemens, embedded_storage_quan}. 

Thus motivated, this paper presents a relatively newer perspective on utilizing energy storage in bulk power systems which is different from their traditional operation as ancillary reserves. The storage resources in this work, are operated as frequency and voltage-responsive dynamic assets which participate in managing transient disturbance events. We envision a network of these storage assets embedded in the power system, and autonomously controlled using local measurements. Their operation, in principle, is similar to fast-acting reserves offering grid reliability functions at inertial and FFR timescales. However, instead of deploying a few large-sized centralized storage units, this work leverages a network of multiple modest-sized grid-scale storage assets distributed across the grid \cite{cdc_gfm_damping_2}. The distribution ensures that the disturbances originating from different parts of a grid are effectively managed and spatially contained in the close neighborhood of their source. The storage resources in the network are \emph{firm designable} $-$ i.e., the utilities have control over the placement and sizing of the assets, and  \emph{firm dispatchable} $-$ i.e., the utilities are also able to decide the operating modes of these assets to meet the objectives of security and resilience. Ideally, these assets are placed at the edge of the transmission (T) network at the substations interfacing with the distribution (D) system. Operating at the T\&D boundary allows the storage network to act as a buffer between the systems to absorb disturbance transients from either side \cite{pnnl_embedded_storage}. 

In this work, the storage resources are interfaced with the power network using droop-controlled grid-forming inverters (GFMs) \cite{REGFM_A1}. GFMs actively control the frequency and voltage at their point of interconnection (POI) and can emulate the dynamic behavior of synchronous generators \cite{weiDu_gfm_gfl, gfm_lasseter}. These inverter controls enable active power modulation from the storage resources to arrest large-scale frequency transients during disturbances \cite{gfm_review}. An added layer of safety-constrained control \cite{kundu2020transient, slac3r} in the GFMs ensures that the frequency nadirs are restricted within the operating boundaries to avoid triggering under/over-frequency tripping during disturbances. 

The following are the contributions of this paper. (1) Using the results from an IEEE test system, we demonstrate that a network of grid-forming storage resources embedded in the bulk power grid, while providing fast-frequency support, can limit the spatial propagation of a frequency disturbance by containing its severity in the proximity of the source. (2) For nodes close to the disturbance source experiencing large frequency deviations, we demonstrate that a distributed safety control local to the inverters can enforce the trajectory of the system frequency to be within a prescribed tolerance band. (3) Sensitivity studies are performed to evaluate the impacts of storage capacity and GFM controller parameters on frequency control and disturbance localization. (4) A comparison between grid-following and grid-forming storage networks is also discussed to highlight the relative merits of the latter.

The remainder of the paper is organized into five sections presenting the system model, problem description and objectives, case studies, and conclusions. 

\section{System Description}
Consider a $n$-bus transmission network with $n_g$ synchronous generators (SGs) and $n_s$ GFM-interfaced storage resources such that, $n_g\,+\,n_s <\, n$.  Each line in the transmission network is represented by its $\Pi$-equivalent algebraic model.


\subsection{Synchronous Generators}
Each synchronous generator is described by a third-order differential equation representing the flux-decay model. Additionally, each generator is also equipped with a first-order speed governor and a manual excitation system. The dynamics of each generator-governor system, ${G}_i$ for $k \, \in\, \mathcal{N}_g\,:\,\{1,\,2\,\dots, n_g\}$, is given by \cite{pai}
\begin{subequations} \label{eq:SG}
    \begin{align}
        \dot{\delta}_k \, &= \, \omega_k \, - \, \omega_0\\ \nonumber
       M_k \, \dot{\omega}_k \,  &= \,- \, D_k \, (\omega_k - \omega_0) \, +\,P^m_{k} \, - \, E'_{q_k}\,I_q \\&~~~~~~~~~~~~~~~~~~~~~~~~+ (X_q - X'_d)\,I_d\,I_q\\ 
       T'_{do}\,\dot{E}'_{q_k}\, &= \,-\, E'_{q_k} \, -\, (X_d - X'_d)\,I_{d_k} \, + \, E_{fd_k}  \\
     T_{ch}\,\dot{P}_k^m\, &= \,-\,P_{k}^m + \frac{1}{R_{g}}(\omega - \omega_0).
    \end{align}
\end{subequations}
where $\delta_k$ and $\omega_k$ are, respectively, the rotor angle and speed, $E'_{q_k}$ is the $q$-axis induced voltage, $E_{fd_k}$ is the field-axis excitation voltage, $P^m_k$ is the mechanical power, $M_k$ and $D_k$ are, respectively, the inertia and damping constants, $R_g$ is the governor droop, $I_d$ and $I_q$ are the $d$- and $q$-axis currents, $X_d$, $X_q$, and $X_d'$ are, respectively, the $d$- and $q$-axis reactances, and $d$-axis transient reactance, and $T'_{do}$ and $T_{ch}$ are, respectively, the time-constants of field circuit and the speed governor. 

\subsection{Grid-Forming Inverter-Interfaced Storage}
The studies in this paper focus on the disturbance dynamics at the transient timescales in the order of milliseconds to a few seconds. The slower dynamics associated with the charging and discharging of the energy storage is therefore neglected and the state-of-charge is assumed to be constant for the duration of the transient disturbance. However, the dynamics of the power electronic converters interfacing with the storage are considered. GFMs act as the grid interface for the storage resources. The inverter dynamics of each storage resource $j \in \mathcal{N}_i: \{1, \, 2, \, \dots, n_s\}$ is described by \cite{REGFM_A1}
\begin{subequations}
    \begin{align}
        \dot{\delta}_j \, &= \, \omega_j \, - \omega_0\\
        \label{eq:omega_inv}
\tau_j\,\dot{\omega}_j \, &= \, \omega_0 \, - \, \omega_j \, + \, m_{p} \, (P^{{s}}_j \, - \, P_{j})\\
\tau_j\,\dot{V}^e_j \, &= \,  V_j^s \, - \,  V_j \, -\, V^e_j\,+ \, m_{q} \, (Q^{{s}}_j \, - \,  Q_{j})\\ 
\dot{E}_j \, &=\, k_{p} \, \dot{V}^e_j \,+ \, k^{v}_j\, V^e_j
    \end{align}
\end{subequations}
where $\delta_j$ and $\omega_j$ are respectively the voltage angle and frequency of the internal bus, $V_j$ and $E_j$ are respectively the voltage magnitudes of the external and the internal buses, $P_j^s$, $Q_j^s$, and $V_j^s$ are respectively the set points of real power, reactive power, and external voltage, $m_p$ and $m_q$ are respectively the $P-f$ and $Q-V$ droop gains, $k_p$ and $k_v$ are respectively the proportional (P) and integral (I) gains of the voltage controller, $V_j^e$ is an internal state used as the input to the PI controller, and $\tau_j$ is the time-constant of the low-pass measurement filter. 

The inverters have an added layer of distributed safety control that adjusts the set points for frequency control. This is discussed next.

\subsection{Distributed Safety Control at the Inverters}

The safety control \cite{kundu2020transient} at a GFM ensures that the frequency at its POI during disturbance events is constrained in an operator-specified range $\left[\, \underline \omega,\, \overline \omega \,\right]$, with $\underline \omega$ and $\overline \omega$ as its lower and upper limits respectively. This is achieved by adjusting the set point $P_j^{{s}}$ based on the local measurements of $P_j$ and $\omega_j$. To this end, two barrier functions,
\begin{subequations}
\begin{align}
      \underline{h}(\omega) \, = \, \omega \, - \,\underline{\omega} ~~~~~\text{and}~~~~
      \overline{h}(\omega) \, = \, \omega \, - \,\overline{\omega}
\end{align}  
\end{subequations}
are defined. Likewise, the following limits (see (\ref{P_barrier})) on $P_j^{s}$ are also defined, where the performance parameters $\alpha$ is a positive value and $q$ is a positive odd integer. 
\begin{subequations} \label{P_barrier}
    \begin{align}
        \overline{P}_j^{s} (\omega) \, &= \, P_j\,+\, \frac{1}{m_p}\,\Big( \omega \, - \, \omega_0 \, -\, \alpha\,\left[\overline h(\omega)\right]^p \Big)\\
         \underline{P}_j^{s} (\omega) \, &= \, P_j\,+\, \frac{1}{m_p}\,\Big( \omega \, - \, \omega_0 \, -\, \alpha\,\left[\underline h(\omega)\right]^p \Big)
    \end{align}
\end{subequations}
The rationale behind choosing these limits will become obvious from the working of the control, explained next. 

When the lower limit on $\omega$ is violated, i.e., $\omega_j \, < \, \underline{\omega}$, to ensure that $\omega_j$ is brought within the safe operating range, the control tries to achieve $\dot\omega_j > 0$.  To do so, the set point $P_j^s$ is updated such that $P_j^s \geqslant \underline{P}_j^{s} (\omega_j)$. This can be explained from (\ref{eq:omega_inv}). Observe that, 
\begin{equation} \label{eq:omega_barrier_exp}
\begin{aligned}
     P_j^s \geqslant \underline{P}_j^{s} (\omega_j) \\ \, \implies \, 
     \frac{\tau}{m_p}\, \dot \omega_j \,&\geqslant \, \frac{1}{m_p}\,(\omega_0 \,- \,\omega_j)\, + \, \underline{P}_j^{s} (\omega_j) \, - \,P_j\\
     \implies \, \frac{\tau}{m_p}\, \dot \omega_j \,&\geqslant \, -\, \alpha\,\left[\underline h(\omega_j)\right]^p .
\end{aligned}
\end{equation}
Since, $\underline{h}(\omega_j) < 0$ for $\omega_j \, < \, \underline{\omega}$, and $\tau, \alpha,$ and $m_p > 0$, and $p$ is positive odd, it follows from (\ref{eq:omega_barrier_exp}) that 
\begin{equation}
    P_j^s \, \geqslant \, \underline{P}_j^{s} (\omega_j) \, \implies \, \dot \omega_j > 0.
\end{equation}
The control objective is thus achieved. Similarly, when the upper limit is violated, i.e., $\omega_j \, > \, \overline{\omega}$, the choice  $P_j^s \leqslant \overline{P}_j^{s} (\omega_j)$ ensures $\dot \omega_j < 0$, and the frequency is steered into the safety range. Once the frequency is in the safe region, i.e., $\underline \omega \leqslant \omega_j \leqslant \overline \omega$, $P_j^{s}$ is re-adjusted to its preset dispatch, say $P_j^{*}$.

Combining these, the control law can be summarized as
\begin{equation}
    \begin{aligned}\label{safcon}
    P^{s}_j = \begin{cases}
        P^{*}_{j} & \; \mbox{if }\; \underline\omega \leqslant \omega_i \leqslant \overline\omega, \\
        \min \big( \,\overline{P}_j^{s}, \, \max \big(\,\underline{P}_j^{s}, \, P_j^{*}\big) \big) & \; \mbox{otherwise}.
     \end{cases}
\end{aligned}
\end{equation}
Note, $P^s_j$ is also constrained by the available power headroom.

With the system model and the control described, we are now well-poised to present the problem description, scope, and objectives of this work. 
\section{Problem Description and Objectives}
The paper addresses the following in the context of frequency control. 

(1) Given a disturbance, resulting from a contingency like a generator trip or any in-feed loss in general, the target is to arrest the frequency transients well within the under-frequency load-shedding (UFLS) limits. Typically, a large part of this support will be contributed by the rotational inertia of the synchronous generators and their governor PFR. However, in low-inertia systems resulting from the retirement of generators, these contributions may not be sufficient. Additional frequency support would be needed from inverter-based resources. To this end, the goal is to demonstrate that a network of grid-forming storage resources distributed in the system can proactively contribute to frequency control during disturbances.  

(2) Secondly, it is also imperative that the severity of a disturbance event be contained in the electrical neighborhood of the source such that its propagation into the wider network is limited. The goal, therefore, is also to demonstrate that a network of modest-sized storage resources dispersed in the power network can achieve this. 

(3) The limits $\overline{\omega}$ and $\underline{\omega}$ for the safety-constrained frequency control are set at $60.2$ and $59.8$ Hz respectively. The UFLS settings in North America, typically for large systems like the Eastern and Western Interconnections, are at $59.5$ Hz \cite{bal-03}. The limits on the safety control, therefore, set a stricter bound on the frequency and allow for a margin with respect to the UFLS. Our goal is to demonstrate that with adequate headroom on the storage resources, these reliability obligations are met.

Case studies demonstrating these are presented next. 

\section{Case Studies and Simulation Results}

\begin{figure*}[t]
    \centering
    \includegraphics[width=\linewidth]{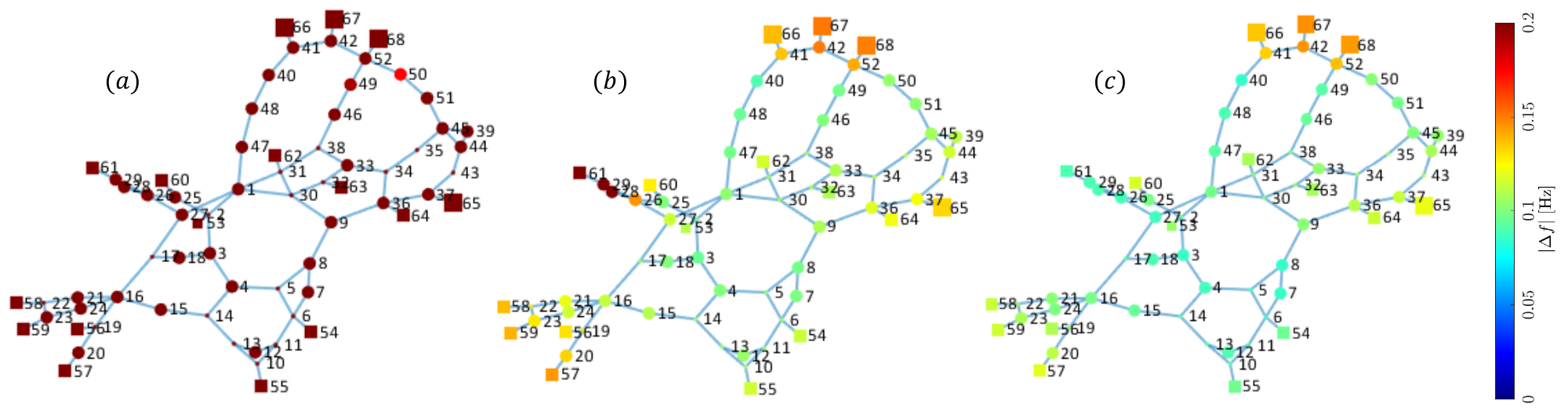}
    \caption{Heat map of $\Delta f$ and spatial propagation of frequency disturbance for G9 generator trip in the IEEE 68-bus (a) base case with no storage, $(b)$ modified case with GFM storage, and $(c)$ modified case with GFM storage and safety control.}
    \vspace{-0.2cm}
    \label{fig:spatial_plot}
\end{figure*}

We consider the positive-sequence simulation model of the IEEE 68-bus transmission system for our case studies. The system consists of a network of $16$ synchronous generators, each represented by a third-order differential equation model describing the electromechanical dynamics of the rotor swings and the field flux. In addition, each generator is also equipped with a speed governor and a manual excitation system. The base case of the 68-bus system is modified to include a network of inverter-interfaced storage resources embedded in the transmission system. The storage assets are co-located with the loads at the substations interfacing the transmission system with aggregated distribution nodes. Droop-controlled REGFM\_A1 model \cite{REGFM_A1} of the GFMs acts as the grid interface for these storage resources. In addition to the $P$-$f$ and $Q$-$V$ droop functions, the GFM model is also equipped with $P$ and $Q$ limiters, and a transient over-current (OC) limiter.

Without loss of generality, the steady-state dispatch $P_j^*$ and $Q_j^*$, in this work, are set to zero. This lets the storage resources operate with optimal headroom in either direction for a given capacity, which allows for deriving maximal frequency support during transient disturbances. 

\subsection{Role of Storage Network in Containing Propagation of Disturbance Events}
We demonstrate how a network of grid-forming storage resources participates in arresting the spatial propagation of a frequency disturbance in a power network. First, consider the base case of the IEEE 68-bus test system with only synchronous generators and without any inverter-interfaced storage resources. 
Generator G9 connected to bus 61 is tripped, and the time evolution of system frequency following the disturbance is studied. A heatmap of $\Delta f$, defined as
\begin{equation}
    \Delta f = \abs{ f_{
    \text{nominal}} - f_{\text{min}}
    },
\end{equation}
is plotted for each node in the power network (see, Fig. \ref{fig:spatial_plot}(a)). Here, $f_\text{min}$ is the minimum frequency or the nadir attained over the disturbance horizon and $f_\text{nominal}$ is the pre-disturbance nominal system frequency. The color spectrum for the heat maps is normalized for the dispersion of $\Delta f$ in the range $0-0.2$ Hz. Further, the nodes with $\Delta f > 0.2$ Hz are uniformly represented by the maximum of the color spectrum. Observe from Fig. \ref{fig:spatial_plot}(a) the frequency transients from the tripping of G9 propagates throughout the power network with $\Delta f$ exceeding $0.2$ Hz at all nodes. 

\begin{figure}
    \centering
    \includegraphics[width=\linewidth]{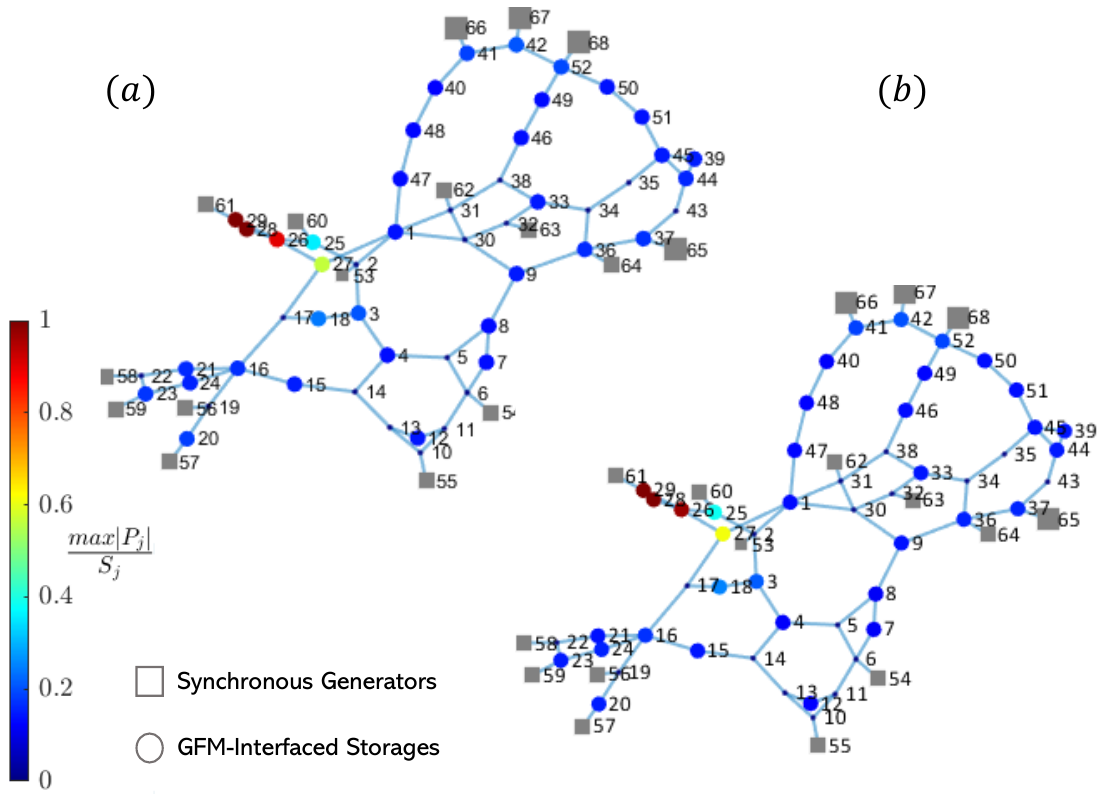}
    \caption{Heatmap of real power injection from the storage units relative to their capacities, $(a)$ without and $(b)$ with safety control.}
    \vspace{-0.5cm}
    \label{fig:pinj}
\end{figure}

Next, we introduce a network of droop-controlled GFM storage resources in the IEEE 68-bus system. The case study is designed such that the total active power capacity of the storage assets is $15\%$ of the total real power load of the system. The capacity is distributed equally among all 35 storage units in the system. For comparison with the base case, the same disturbance event is simulated, i.e., G9 is tripped, and the $\Delta f$ heatmap is plotted in Fig. \ref{fig:spatial_plot}(b). Observe that compared to Fig. \ref{fig:spatial_plot}(a), the frequency nadirs in Fig. \ref{fig:spatial_plot}(b) are improved for most of the nodes in the system. Also, the severity of the disturbance in Fig. \ref{fig:spatial_plot}(b) is localized to the buses $29, 28$, and $26$ which are adjacent to the source of the disturbance at bus $61$. The plot in Fig. \ref{fig:pinj}(a) shows the heat map of the maximum real power injections from the storage resources. The units closer to the point of the disturbance have a larger contribution to the power injection. Even with the maximum contributions from the storage resources at buses $28$ and $29$ the frequency nadirs at those buses are not appreciably improved. Although the stress on the remaining system is greatly relieved, these nodes still have $\Delta f \geqslant 0.2$ Hz. This is due to the complete utilization of the capacities of the local storage units leaving no headroom for providing additional frequency support. 

An easy fix to alleviate frequency nadirs at the buses close to the source of the disturbance is to increase the active power capacity of the storage units. However, that can be highly cost-prohibitive as the increased capacity would remain under-utilized for all instances other than large disturbances. Instead, we propose to retrofit the storage units with the safety control, discussed in Section II, which ensures frequency nadirs do not exceed the $0.2$ Hz limit. This is presented next.

\subsection{Retrofitted Safety Control to Contain Frequency Nadirs Within Operating Limits}
\begin{figure}
    \centering
    \includegraphics[width=0.49\linewidth]{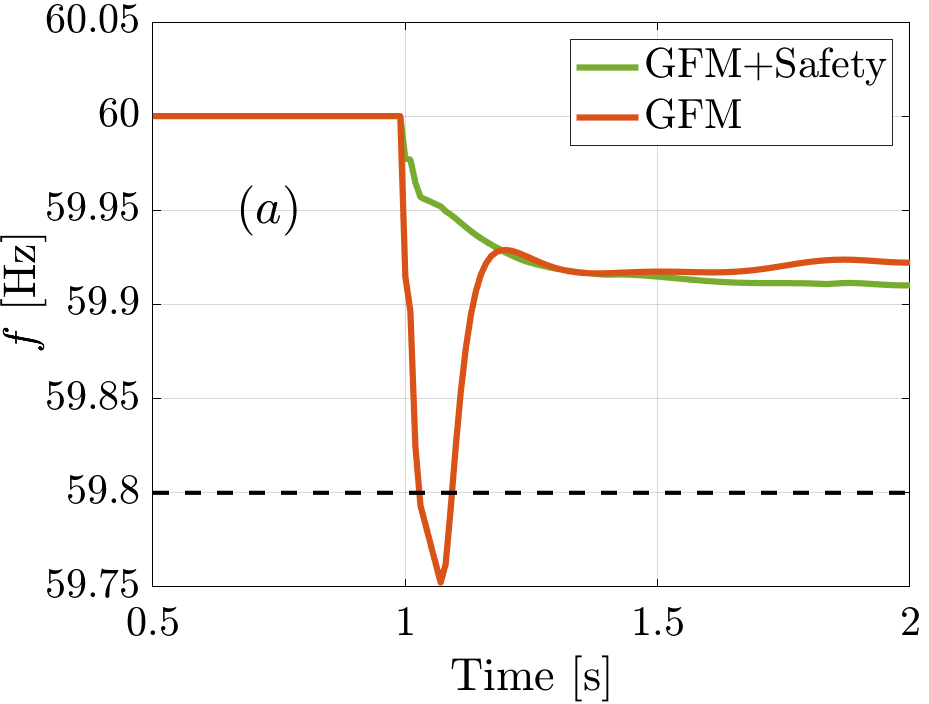}
    \includegraphics[width=0.49\linewidth]{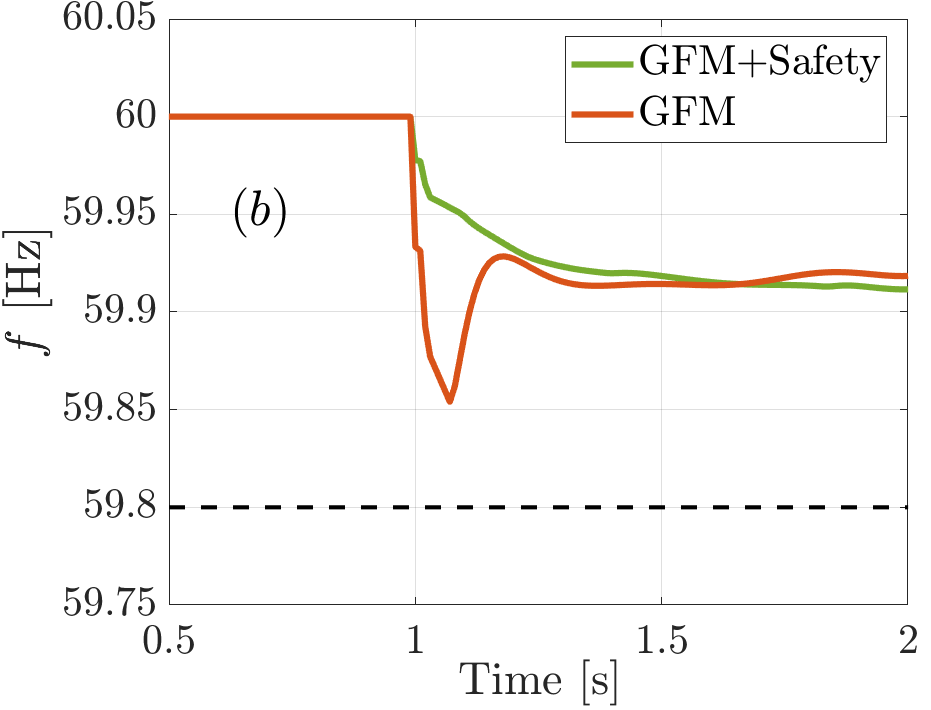}
    \caption{Frequency profile at the load buses $(a)$ 26 and $(b)$ 29, with and without safety control on the GFM storage network.}
    \label{fig:freq_profile}
    \vspace{-0.5cm}
\end{figure}

\begin{figure}[h]
    \centering
    \includegraphics[width=0.49\linewidth]{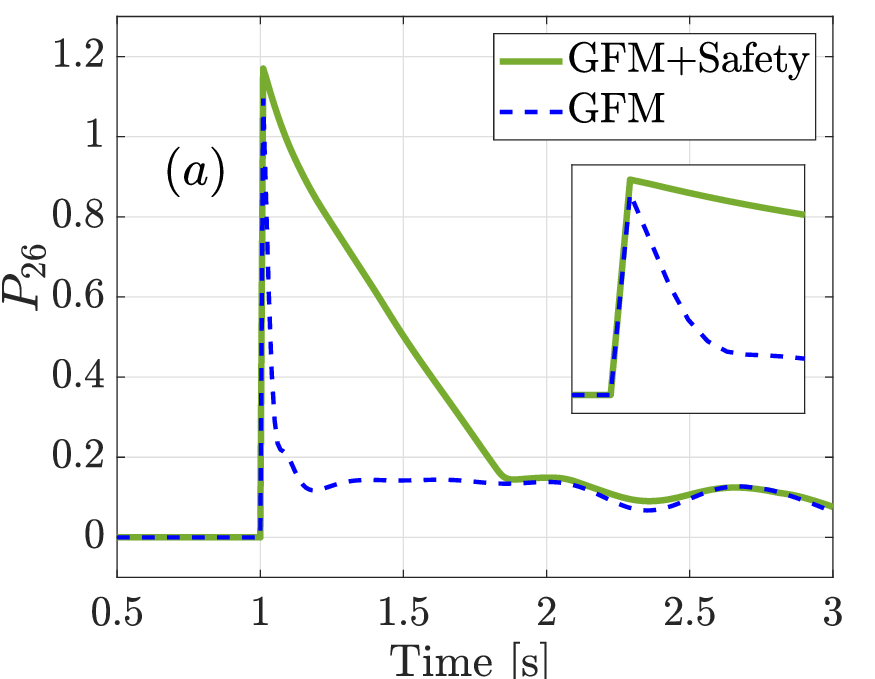}
    \includegraphics[width=0.49\linewidth]{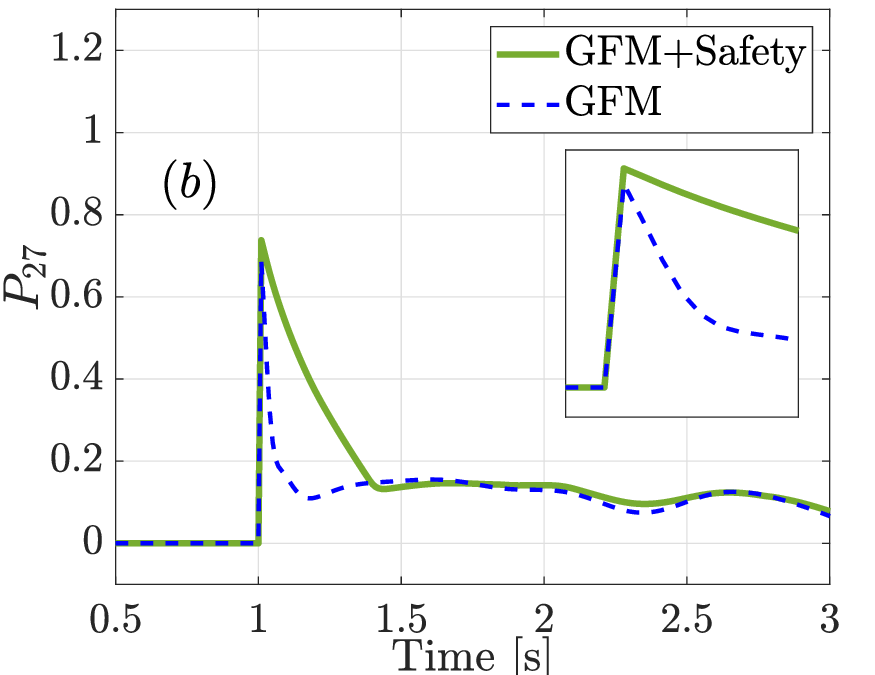}
    \includegraphics[width=0.49\linewidth]{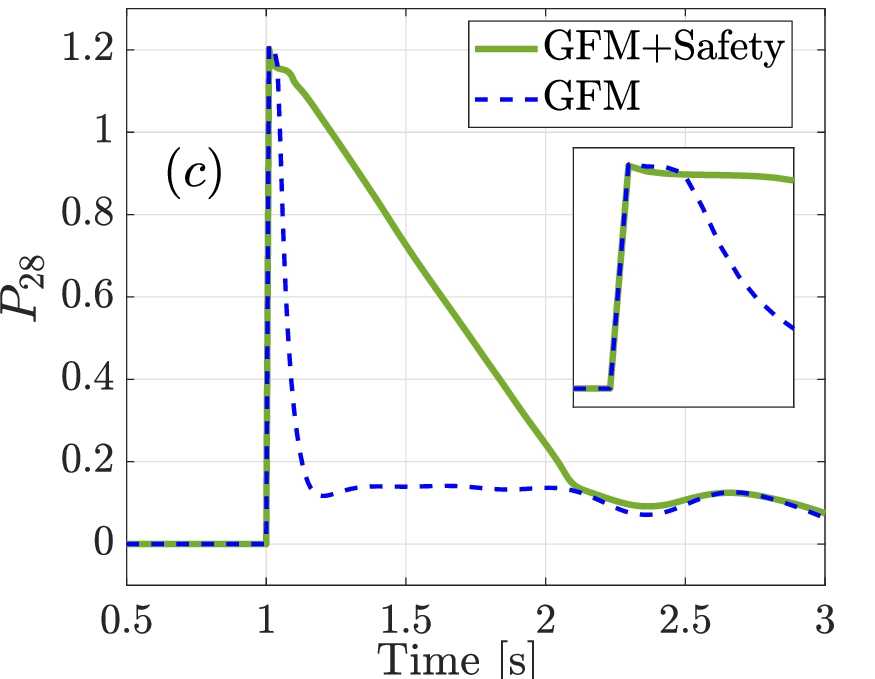}
    \includegraphics[width=0.49\linewidth]{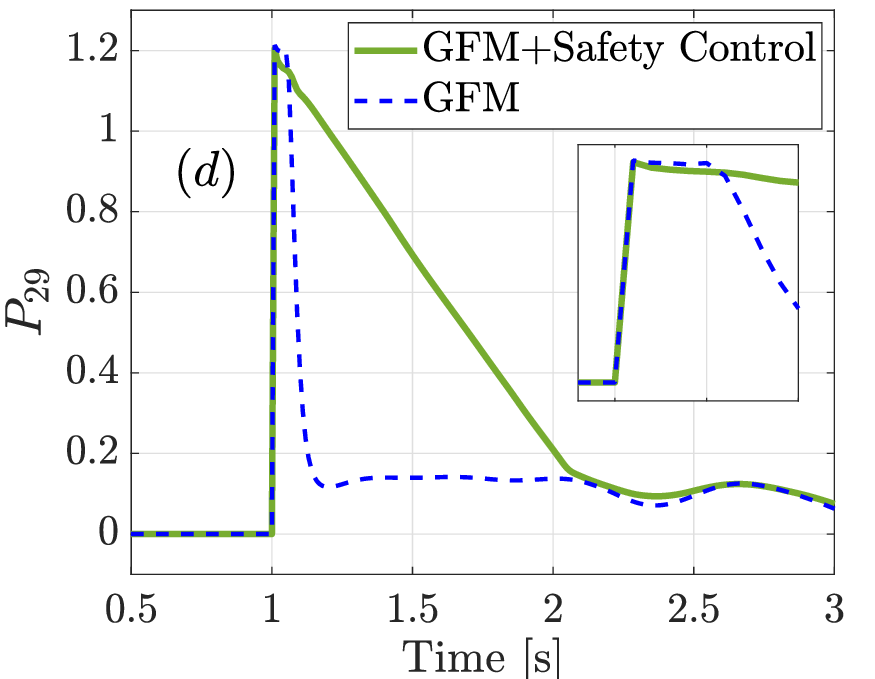}
    \caption{Relative real power injections from the GFM storage units connected to buses $(a)$ 26, $(b)$ 27, $(c)$ 28, and $(d)$ 29, with and without safety control.}
    \label{fig:15GFM_inj}
    \vspace{-0.2cm}
\end{figure}

In this, the real and the reactive power setpoints of the GFMs interfacing with the storage resources are dynamically updated based on a decentralized control law computed from the local measurements of bus frequency, voltage, and output power. The control law utilizes barrier functions that are activated as the performance variables (frequency and voltage) approach the boundaries of the safety limits. The upper and lower safety limits for the frequency, as mentioned, are set to $60.2$ and $59.8$ Hz respectively.  The details of the control implementation are outlined in Section II. Same as the cases discussed before, G9 is tripped and the heatmap of $\Delta f$ is plotted in Fig \ref{fig:spatial_plot}(c). Observe, that the introduction of the local safety control improves the frequency profile in the neighborhood of the disturbance and contains the $\Delta f$ at the buses $28, 29$, and $61$ within the $0.2$ Hz limit. The frequency profiles at buses $26$ and $29$ with and without the safety control are shown in Fig. \ref{fig:freq_profile}. Further, as seen from the heatmap of inverter maximum power injections in Fig. \ref{fig:pinj} (b), the contributions from the storage resources at buses $26$ and $27$ are slightly higher compared to the case in Fig. \ref{fig:pinj}(a). The power injections from these storage resources for the two cases, with and without safety control, are also compared in the time domain plots of Fig. \ref{fig:15GFM_inj}. The increase in power injections for the latter is due to the safety-control which adjusts the real power setpoints at these GFMs to maximally utilize their headroom and arrest the excursion of system frequency outside the prescribed limits. 

\subsection{Impacts of Storage Capacity and Controller Gain}
\begin{figure}
    \centering
    \includegraphics[width=0.49\linewidth]{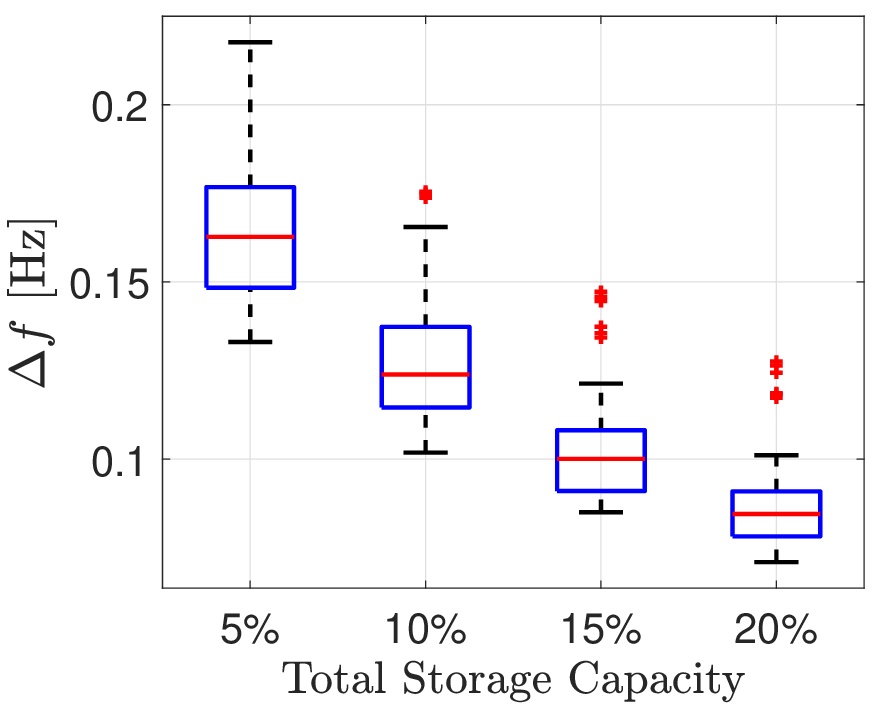}
    \includegraphics[width=0.49\linewidth]{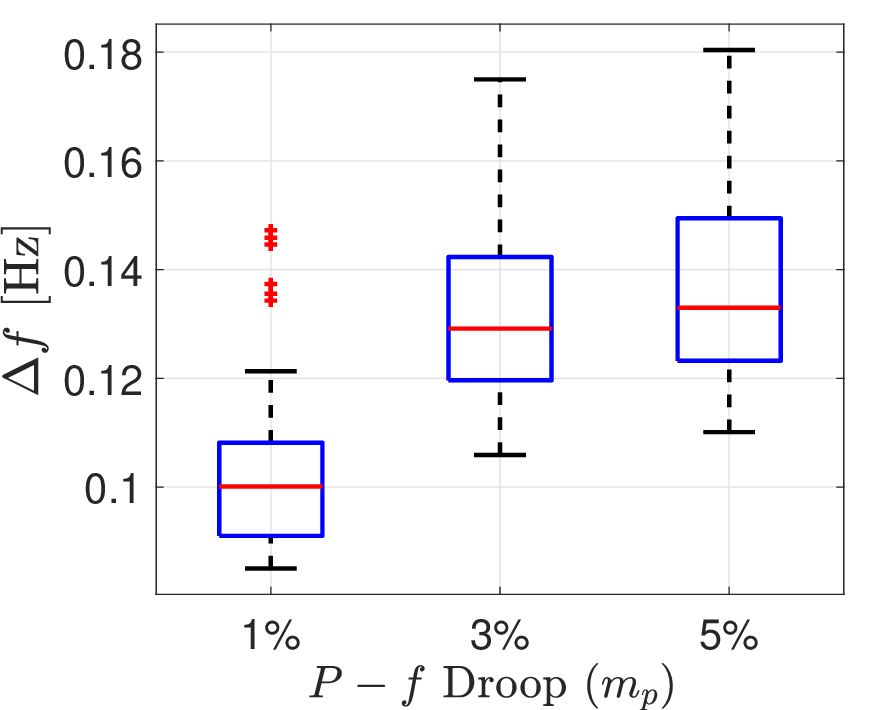}
    \caption{Box plots of $\Delta f$ for all 68 buses showing the impacts of $(a)$ storage capacity (expressed as $\%$ of total load) and $(b)$ GFM real power-frequency ($P-f$) droop gain, in the presence of safety control.} 
    \label{fig:freq_boxplot}
    \vspace{-0.4cm}
\end{figure}

Next, we study the impacts of storage capacity and GFM droop gain on the performance of the frequency control. We vary the cumulative capacity of the storage assets from $5\%$ to $20\%$ of the total real power load in the system. For the same G9 trip disturbance event, we see that the median $\Delta f$ reduces with an increase in the storage capacities (see, Fig. \ref{fig:freq_boxplot}(a)). This is intuitive since a larger headroom allows for more efficient frequency control. 
The impact of GFM real power-frequency droop gain $m_p$ on the frequency control is also studied. As seen in Fig. \ref{fig:freq_boxplot}(b), for a fixed storage capacity, lower values of $m_p$ lead to well-damped frequency transients and better-regulated frequency nadirs compared to a higher droop case. 

\subsection{Comparison with Grid-Following Storage Resources}
\begin{figure}[h]
    \centering \vspace{-0.65cm}
    \includegraphics[width=0.85\linewidth]{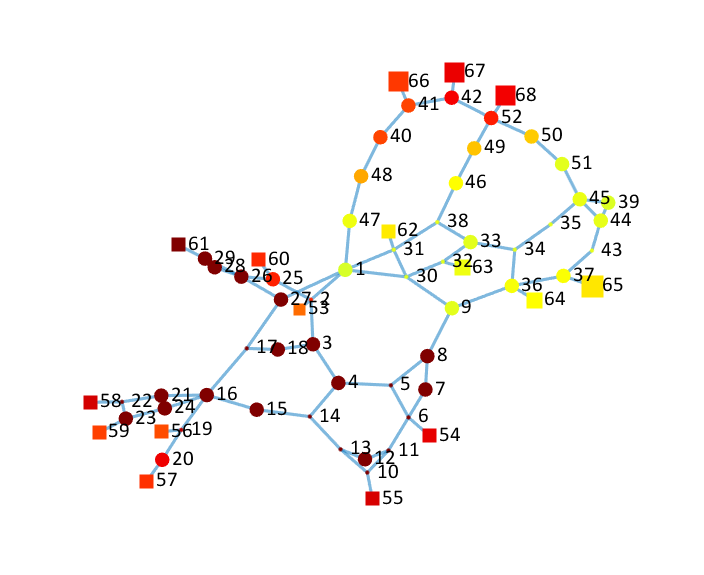}\vspace{-0.7cm}
    \caption{Heat map of $\Delta f$ and spatial propagation of frequency disturbance for G9 generator trip in the IEEE 68-bus modified with GFL storage network. }
    \label{fig:gfl_heatmap}
    \vspace{-0.4cm}
\end{figure}

Finally, we compare the results of disturbance containment using the grid-forming storage with their grid-following counterparts. For this study, the GFMs interfacing with the storage resources are replaced with a generic model of the grid-following inverters (GFLs) \cite{weiDu_gfm_gfl}. The heatmap of $\Delta f$ for the G9 trip with a GFL storage network is shown in Fig. \ref{fig:gfl_heatmap}. The frequency deviations in Fig. \ref{fig:gfl_heatmap} are larger compared to those in Fig. \ref{fig:spatial_plot}(b) and (c). Also, the severity of the disturbance percolates throughout the network unlike the GFM case, where it was contained in the electrical neighborhood of the source.  

\section{Conclusions}
The paper studied grid-forming storage networks embedded in bulk power grids for sourcing frequency support during transient disturbances. It was shown that a network of grid-forming storage resources operating with adequate real power headroom contains the severity of a frequency event in the neighborhood of the disturbance source, thereby limiting its spread to other regions of the network. Further, a safety-constrained local control was discussed, which when added to the existing GFM controls, can improve the frequency nadirs of the worst-hit nodes by restricting their frequency transients within a preset safe operating band. The paper also presented results from sensitivity studies to explore the impacts of storage size and GFM droop gain on frequency control. It was also demonstrated from case studies, that the GFM-interfaced storage networks are more effective in frequency control than the GFL networks.   

\bibliographystyle{IEEEtran}
\bibliography{Ref_damp}

\end{document}